\pdfoutput=1
\documentclass[letterpaper,twocolumn,10pt]{article}
\usepackage{usenix2019_v3}

\usepackage{tikz}
\usepackage{amsmath}
\PassOptionsToPackage{hyphens}{url}
\usepackage{xurl}
\usepackage{hyperref, graphicx}
\graphicspath{{images/}}
\usepackage{comment}
\usepackage[utf8]{inputenc}
\usepackage[T1]{fontenc}
\usepackage[french,english]{babel}
\usepackage{verbatim}
\usepackage[ruled,vlined]{algorithm2e}
\usepackage{amsmath, amsfonts}
\usepackage{subcaption}
\usepackage{multirow}

\newcommand{\aname}{ACT}
\newcommand{\company}{eBay}
\newcommand{\heading}[1]{\vspace{1mm}\noindent\textbf{#1}\hspace{1mm}}

\usepackage[skip=2pt]{caption}
\usepackage[aboveskip=2pt]{subcaption}
\usepackage[compact]{titlesec}
\newcommand{\Figure}{Fig.}
\makeatletter
\renewcommand{\fnum@figure}{Fig. \thefigure}
\makeatother

\begin{document}

\date{}

\title{\Large \bf ACT now: Aggregate Comparison of Traces for Incident Localization}

\author{
{\rm Kamala Ramasubramanian}\\
University of California, Santa Cruz
\and
{\rm Ashutosh Raina}\\
eBay
\and
{\rm Jonathan Mace}\\
MPI-SWS
\and
{\rm Peter Alvaro}\\
University of California, Santa Cruz
}

\maketitle

\begin{abstract}
Incidents in production systems are common and downtime is expensive. Applying an appropriate mitigating action quickly, such as changing a specific firewall rule, reverting a change, or diverting traffic to a different availability zone, saves money. Incident localization is time-consuming since a single failure can have many effects, extending far from the site of failure. Knowing how different system events relate to each other is necessary to quickly identify \emph{where} to mitigate. Our approach, Aggregate Comparison of Traces (\aname), localizes incidents by comparing sets of traces (which capture events and their relationships for individual requests) sampled from the most recent steady-state operation and during an incident. In our quantitative experiments, we show that \aname\ is able to effectively localize more than 99\% of incidents.
\end{abstract}

\section{Introduction}

Users expect web services to be highly available. Outages of even a few minutes can cost service providers hundreds of thousands of dollars in revenue~\cite{GoogleOutage2013, AmazonOutage2013}. Additionally, they incur soft costs in terms of poor user experience. Site Reliability Engineers (SREs) declare an incident when system functionality becomes unavailable. Before they can act to mitigate system unavailability, SREs must first \emph{localize}  the incident. Localization is the process of identifying a location - a component (hardware or software) - where a mitigating action may be applied. For example, if SREs determine that the behavior of service instances in a particular data center is problematic, the action recommended by SREs might be to divert traffic away from it. Other examples of mitigating actions include re-configuring access control or firewall rules, reverting a code or configuration change, restarting components, etc. Mitigating system unavailability, when it occurs, is the priority for SREs.

At first glance, it seems like debugging~\cite{ZhangSOSP2019, WhitakerSoCC2018, ZhangSOSP2017, ScottNSDI2016} and software fault localization~\cite{SahooASPLOS2013,B.LeISSTA2016} techniques may be applicable to incident localization. However, neither the proposed approaches nor the time scales at which they operate are applicable. These techniques focus on root cause analysis to help developers identify a software or hardware fix to the underlying problem. As an example, assume that a recent change is causing service instances to crash and restart repeatedly. In this case, localization to the service in question is sufficient for SREs to mitigate the incident by identifying and reverting the change, \emph{even if they don't understand what went wrong!} Debugging or software fault localization techniques, on the other hand, focus on finding a line of code or a parameter to be adjusted to fix system functionality. Incident localization is also much more time constrained as compared with root cause analysis, a task undertaken once the system is available again. Therefore, debugging and software fault localization are complementary to incident localization.

Incident localization is difficult in practice for two reasons. First, since real-world distributed applications are complex and highly connected, SREs need to consider large volumes of data from varied sources (metrics, logs, events, and traces) generated by executions before and during the incident to reason about system behavior. Based on their observations, SREs then attempt to determine a pattern in \emph{how} executions fail during an incident. Second, many events in the failing executions may be different from the successful executions. SREs have to infer the relationships between different events for effective (correct and precise) localization; a time-consuming process. SREs have access to a suite of tools but may need to use multiple tools to obtain insights from different data sources. Outputs from one tool may be modified and used as inputs to a different tool.  
As the time to localize an incident increases, so does the time taken to mitigate unavailability, costing money.

For each user request, a trace captures events within a user request and how they relate to each other. The relationships captured in traces are precisely those which SREs currently infer manually recommending the use of traces. Comparing sets of traces helps determine the consistent structural changes across traces (change pattern) during an incident. Thus, using traces as a data source addresses both causes of slow incident localization and automates the process.

The key idea of our approach, Aggregate Comparison of Traces (ACT), is to find events present in one set of traces but not the other and then use the structure of individual traces to reason about cause and effect. ACT leverages the relationships captured between events within traces as opposed to SREs manually connecting the dots. Thus, we are able to focus SREs attention on a few (ideally one) events or relationships that they need to investigate further to take an effective mitigating action. 

We evaluate ACT on datasets from HDFS~\cite{ShvachkoMSST2010}, DeathStarBench \cite{GanASPLOS2019}, and \company. In our quantitative experiments, we conduct hundreds of simulations for three different failure modes and show that ACT is able to identify a mitigation site that enables effective action in all but a handful of cases as compared with our baselines that produce irrelevant results in 30-50\% of the cases. For SREs mitigating incidents, the above result implies that ACT identifies exactly where the mitigation is to be applied. We have integrated ACT with Jaeger~\cite{Jaeger}, an open source tracing tool, for online trace comparison. \aname\ opens a line of inquiry into using groups of traces for incident localization, which, if adopted widely, can change the way SREs approach incident response. 
We also contrast \aname\ with approaches taken by commercial tools such as Lightstep \cite{Lightstep}.

The rest of the paper is organized as follows: In Section \ref{bkground_motivation}, we first present details of an incident study that justifies our choice to focus on incidents which produce structural changes in traces. With an example incident, we motivate the use of traces to localize incidents by discussing relevant approaches from the state of the art and highlighting their shortcomings. Section \ref{design} develops an approach that compares sets of traces and analyzes their events and relationships to localize incidents. In Section \ref{eval}, we focus on evaluating ACT vis-a-vis baselines that adapt approaches from prior work to our setting and finally, we touch upon the details of integrating ACT with Jaeger.

\section{Background and Motivation}
\label{bkground_motivation}

The principal goal of incident mitigation is to minimize impact to users. Understanding the causes of the incident is usually a secondary goal, often a more costly (in terms of time and effort) exercise reserved for post-incident reviews.

SREs use aggregate alerts from metrics deviations, logs from services, etc to build a mental model of the system. These models are often based on tribal knowledge, typically incomplete and usually outdated~\cite{GraysonQueue2020, ReedQueue2020, CookQueue2020}. We first present observations from a study of incidents at \company.

\subsection{Incident Study}
\begin{figure}[t!]
\centering
\includegraphics[width=0.75\columnwidth]{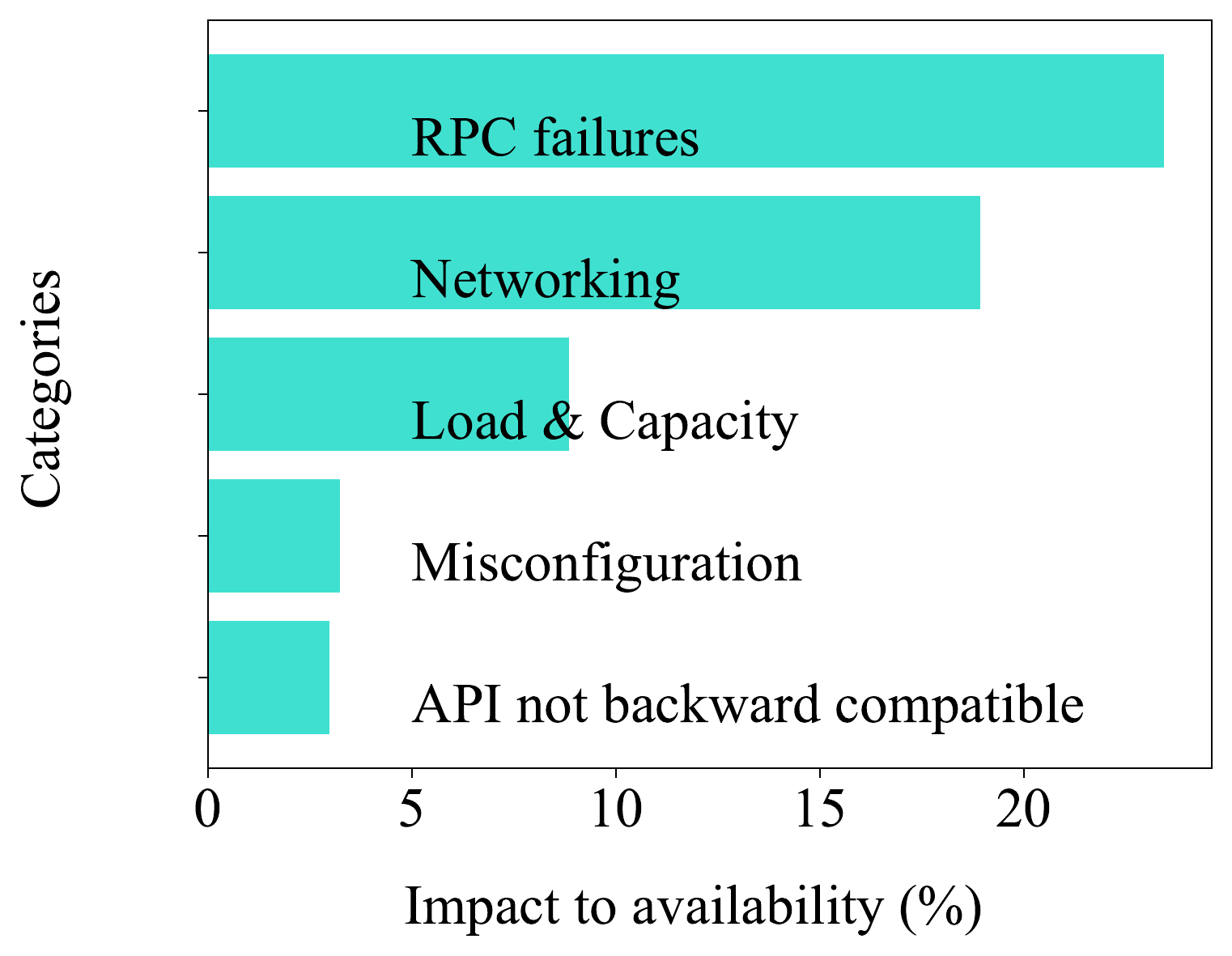}
\caption{Percentage of impact by category - we have represented the top five of over a dozen different categories that emerged based on available data. Incidents arising due to breakdown in communication between components at the application level have the highest impact.}
\label{category_impact}
\end{figure}
To determine trends in incidents, we studied incident reports of 75+ severe incidents that occurred over three years (June 2017 - May 2020) at \company . Severe incidents correspond to more than 85\% of overall impact. Incident impact is measured in terms of loss of availability. Impact to availability of an incident is the duration of time that all or some fraction of users were unable to use the system i.e. availability of service. We make the following observations:

\noindent \textbf{\textit{Highest impact is from RPC failures between components at the application-level:}} \Figure~\ref{category_impact} shows the top five incident categories in order of decreasing impact. We have not represented incidents arising from vendor issues since reading incident reports only gives us a partial view of these incidents. RPC failures between components includes: 

\textbf{Component  down:} 
Components can fail for various reasons - a recent change made to the component, a dormant bug in a code path not used often - triggered by an increase in load or a change in user options. Component failure by itself is not an incident but the lack of a fallback mechanism or the critical nature of a component failure not being common knowledge can lead to an incident. Incidents corresponding to this category may arise due to component failures or decommissioning components that are in use.

\textbf{Component A unable to call component B:}  
A component (A) which was previously able to make RPC calls to a different component (B) that it depends on may no longer be able to do so due to link failures, changes in access control lists, firewall rules, and security fixes. This may further result in additional, unexpected component interactions.

\textbf{Buggy failure recovery:}
When a component fails to respond within the configured timeout or crashes, the calling component may call a different component or perform a series of actions to recover. Since the recovery path is generally not exercised often, it may not work as expected resulting in overall request failure. In such a case, the fallback is inappropriate or incorrectly set up to recover from the failure. This is an important failure mode as failures in the recovery path continue to be reported~\cite{LiuHotOS2019, CloudBugsSOCC2016, GunawiSOCC2014} despite various efforts to address them.

\begin{figure}[t!]
\centering
\includegraphics[width=\columnwidth]{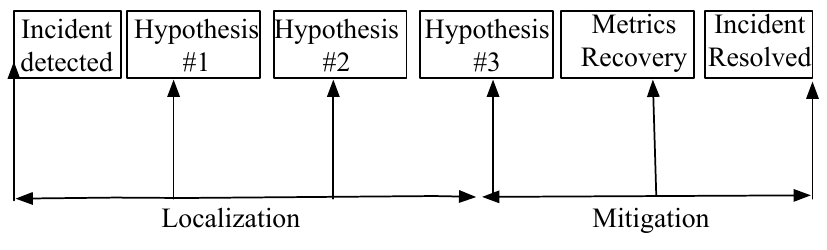}
\caption{This figure represents a typical incident timeline}
\label{incident_timeline}
\end{figure}
\noindent \textbf{\textit{About half the incidents were localized incorrectly at least once:}} \Figure~\ref{incident_timeline} shows a simplified timeline from incident detection to resolution. Localization time and mitigation time are the times taken to effectively localize an incident and apply the mitigating actions that effect recovery respectively. Incorrect localizations (Hypothesis\#1 and Hypothesis\#2) for an incident indicate that there were one or more mitigation steps that were pursued before the incident was effectively localized. A mitigating action that does not result in metrics recovery prolongs poor user experience and increases revenue impact. Prior works~\cite{WangICSE2021, ChenICSE2019, ChenASE2019} indicate that most incidents are reassigned at least once during triage and that triage time dominates response time~\cite{ChenASE2019}. Reassigning incidents results in a longer time to apply a mitigating action and therefore, slower incident response.

From these observations, effective localization of incidents that arise from RPC failures between application-level components would produce the most significant reduction in user impact and therefore, we focus on these.

\subsection{Motivating Example}
\begin{figure*}[t!]
\centering
\includegraphics{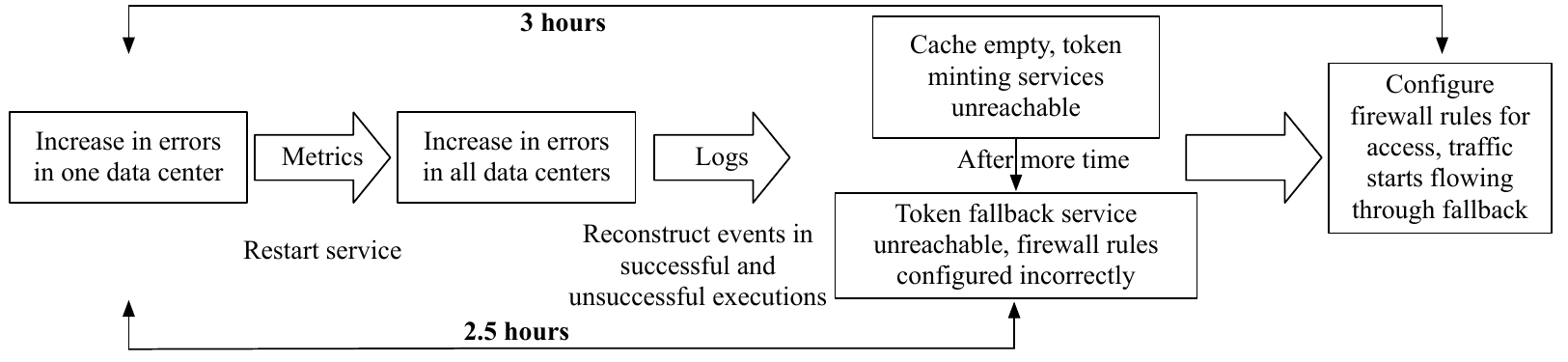}
\caption{This figure represents how SREs responded to an incident and the data sources they used (logs and metrics). The mitigation steps took SREs close to three hours, two and half of which was arriving at the correct mitigating action.}
\label{incident_progression}
\end{figure*}
\begin{figure}[t!]
\centering
\includegraphics[width=\columnwidth]{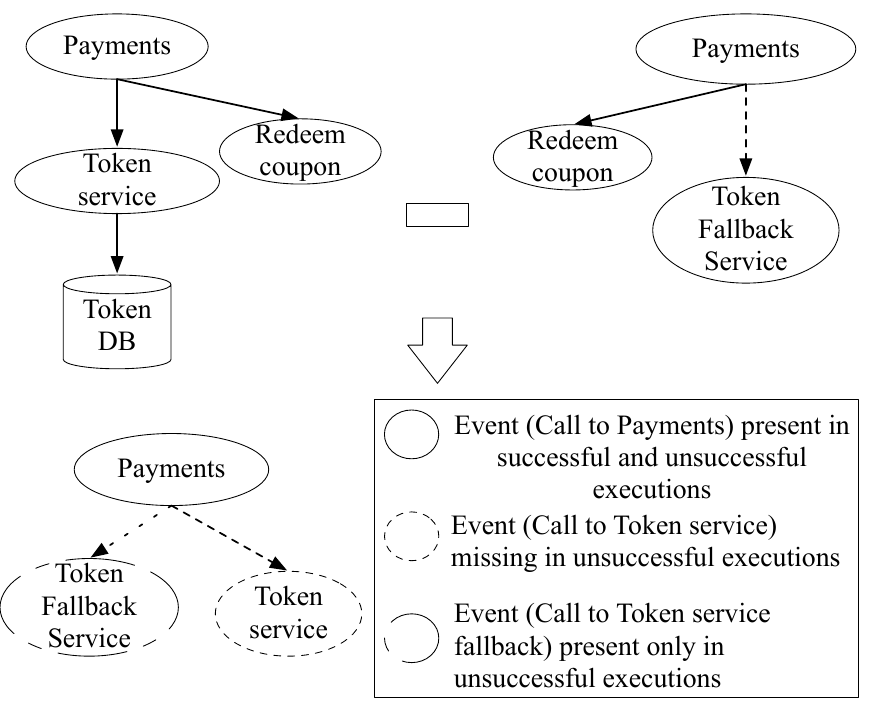}
\caption{This figure represents an idealized picture of graph differencing and contains only the relevant services. On the left is a partial view of a successful request where the token service was working as expected. On the right, we have the trace, after the restart of the payments service which continued to see errors due to incorrect firewall rules.}
\label{Graph_Differencing}
\end{figure}
We describe a real incident that occurred at \company\ which serves as our running example for the remainder of the section. \Figure~\ref{incident_progression} depicts the actions taken by SREs. The mitigation steps took SREs close to 3 hours and was dominated by time taken to arrive at the correct mitigating action (2.5 hours). 

SREs first observed an increase in errors (a metric) for the \textit{Payments Service} in one data center and immediately declared an incident. Metrics are used to monitor the overall health of the system. Business metrics such as number of transactions completed, number of canceled transactions and rate of incoming traffic are tracked in real-time since they are related to revenue. SREs attempted to mitigate the incident by restarting the service, but errors increased in all the data centers instead. In this instance, SREs could surmise from the metrics \textit{that} something was wrong, but not what or why. 

To understand what caused the increase in errors, SREs looked at the application logs and noticed that the local cache used for storing access tokens was empty and the service used for minting the tokens (\textit{Token Service}) was unreachable. SREs found that \textit{Payments Service} started calling \textit{Token Fallback Service} instead of \textit{Token Service}. However, all calls to \textit{Token Fallback Service} were also failing. Further investigation using the logs revealed that the \textit{Token Fallback Service} was inaccessible due to incorrect firewall rules. Once the firewall rules were corrected, the error rate returned to normal and \textit{Payments Service} fully recovered. 

The breakthrough in our running example came when one of the SREs observed \textit{from the logs} that during the incident, requests were attempting to make a call to a service (\textit{Token Fallback Service}). No such call was present in pre-incident execution traces. SREs had to trawl through logs to find the specific events and event interactions in the unsuccessful executions that contributed to its failure. These events indicated the presence of \textit{additional} calls that were not present in executions before the incident. In this case, SREs needed to not only understand the \textit{absence} of calls from the logs but also the \textit{presence} of additional calls. This illustrates that effectively localizing incidents usually requires both aggregate (the presence of errors) and causal information (\textit{Payments Service} trying to call \textit{Token Fallback Service}) - in this instance provided by metrics and logs. SREs had to first determine which executions to consider based on the failure of requests and then compare the events and their relationships between successful and unsuccessful executions. The request path and system model were reconstructed from system logs for this incident. The crucial step in localizing the incident was differencing the request paths before and during the incident to see what was different about the request paths.

\subsection{Limitations of existing approaches}
Although prior work localizing incidents in data centers~\cite{ArzaniNSDI2018, RoyNSDI2017} is extensive, these are orthogonal to localizing application-level incidents since applications are designed to tolerate network failures such as link failures and packet drops. For example, a service usually has instances in multiple data centers such that if a network link in one data center goes out, requests will be sent to a different instance.

We will briefly describe each of the observability signals - metrics, logs, and traces - and the most relevant approaches that use them as inputs. With our running example as context, we discuss why they don't effectively localize incidents. We also discuss the constraints of differential reasoning when using traces and how our approach addresses them.

Fa~\cite{DuanICDE2009} detects and localizes incidents by vectorizing metrics to learn incident signatures. The localization points to the set of metrics (and underlying components) most relevant to the failure. During an incident, multiple metrics are affected and SREs would need to understand relationships amongst different components. Marianil et al.~\cite{MarianiICST2018} use metrics to learn a baseline model and build out an undirected graph by correlating pairs of metrics. They further use graph centrality measures to identify the most severely affected metrics and thereby, a faulty service. For our example incident, the Payments Service has the most errors, and will most likely be identified as the faulty service. This does not give SREs any actionable insights and is therefore, not useful. 

More recent approaches such as Grano~\cite{WangVLDB2019} and Groot~\cite{WangASE2021} assume that relationships amongst components are available either in the form of system architecture diagrams or global dependency graphs. They build machine learning models to identify the metrics correlated with a given incident which are then overlaid on the dependency graph for incident localization. Such follow-the-errors approaches do not work when multiple incidents co-occur, one or more metrics are not captured or incident localization involves identifying when a call between two components \textit{did not occur}. Our example incident falls into this last category.

Logs capture a machine centric view of the system and provide additional context, but require sifting through large volumes of data to extract it. Aggarwal et al.~\cite{IBMFaultLocalization} model logs from different components as multiple time series and correlate errors emitted by various services to localize the incident given a dependency graph (static topology or architecture diagram). Approaches that reconstruct individual user requests from logs involve control and data flow analysis~\cite{ZhaoOSDI2014, YuanASPLOS2010}. Yet others use unique identifiers to identify events corresponding to different requests~\cite{ChowOSDI2014} and custom log parsing to recognize identifiers~\cite{TanHotCloud2009}. Network communication and temporal order are used as heuristics to infer relationship amongst events. Causality inference using log analysis is brittle since it depends on the quality of user logging and is inapplicable either due to practical concerns (running control and data flow analysis for constantly evolving systems like microservices with continuously changing topologies is impractical) or because the timescales for incident localization are very stringent (as systems scale, application logs grow, increasing analysis time).

\textit{Distributed tracing} provides a request-level view of the system and is used for debugging~\cite{BeschastnikhCACM2016, SambasivanNSDI2011, MannHotCloud2011}, profiling, and monitoring production applications. It has also been used to address correctness concerns~\cite{SigelmanReport2010}, for capacity planning, and workload modeling~\cite{BarhamOSDI2004}. Tracing is increasingly being adopted by industry and there is a push for standardization~\cite{OpenTelemetry,OpenTracing, OpenCensus} as well. A trace captures events that occur in a given request as well as how they relate to each other i.e causality. The most general representation of a trace is a directed acyclic graph (DAG) where nodes and edges correspond to events and their interactions respectively.

\begin{figure*}[t!]
\centering
\includegraphics{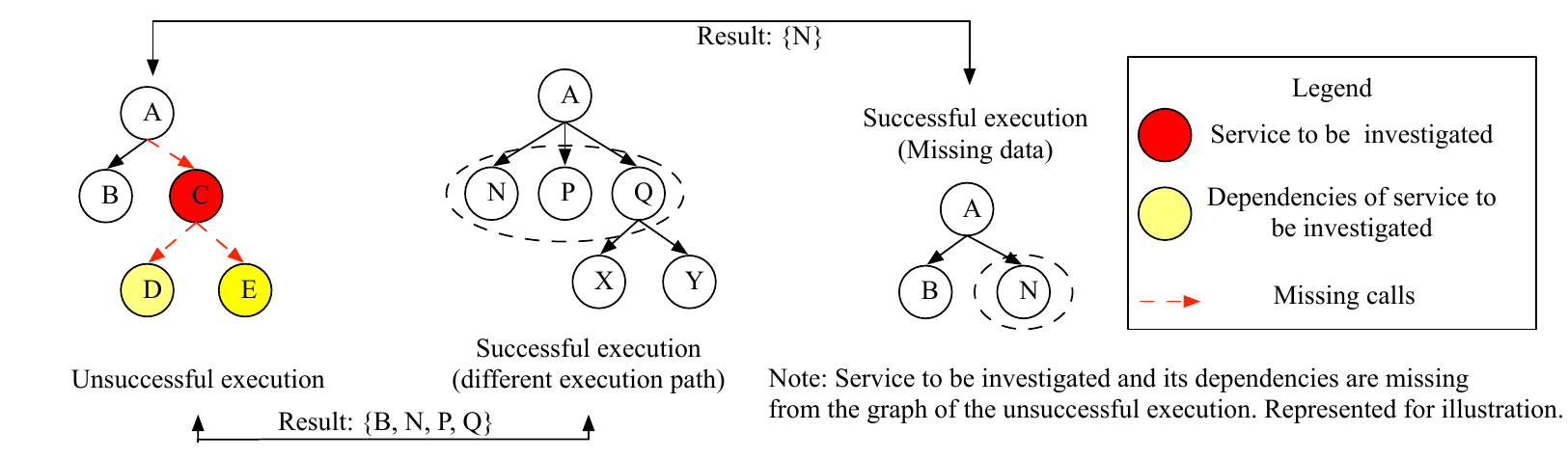}
\caption{Limitations of pairwise comparison - the two examples demonstrate the circumstances when pairwise comparison produces false alarms and can occur either separately or in combination.}
\label{pairwise_comparison}
\end{figure*}
Had traces been available, SREs would been able to compare the trace of a successful execution and that of an unsuccessful execution - which we call \textit{pairwise comparison} - to determine the events that differentiate the two. Doing so would have highlighted the missing and additional events in executions during the incident and thereby enabled them to take effective action. \Figure~\ref{Graph_Differencing} demonstrates an idealized result of pairwise comparison for our running example. In the unsuccessful execution, the call from \textit{Payments Service} to \textit{Token Service} service is missing, but an attempted call from \textit{Payments Service} to \textit{Token Fallback Service} is additional. Since the structure of a trace represents causality of event interactions, we use it to establish cause-and-effect relationships between events in the result - retaining only the causes.

Prior works that uses trace analysis employ a similar differential approach between pairs of traces with appropriate user inputs. For eg., ShiViz~\cite{BeschastnikhCACM2016} and Jaeger~\cite{Jaeger} both support pairwise comparison of user selected traces. Such tools allow SREs to interactively validate hypotheses but are not suited to automated localization. 

\subsubsection{Pairwise Comparison: A deep dive}
The most important requirement for automated localization using pairwise comparison is \textit{graph selection} - selecting a successful and an unsuccessful execution that exercise the same code path. In large, distributed systems, requests with identical inputs can often take different paths due to cache effects, dynamic request routing, traffic shifting across data centers, experimentation, etc. 
Traces generated from such requests may have different structures wholly or partially. 
Further, the structure of traces can also change with configuration changes in the application and deployment environment, ongoing code deployments, new feature deployment and code deprecation. At any given time, several such changes to request paths exist in production. Comparing pairs of graphs corresponding to different executions paths will produce incorrect localizations.

Further exacerbating the problem of graph selection is the fact that tracing is best-effort. That is, for some executions, the trace corresponding to the execution may be missing some data. \Figure~\ref{pairwise_comparison} demonstrates incorrect localizations produced by comparing executions that exercise different execution paths and when comparing incomplete traces of similar executions. Therefore, choosing a pair of graphs to compare based only on their structure is not viable.

Prior work makes simplifying assumptions about the system under consideration to make graph selection viable. Magpie~\cite{BarhamHotOS2003} assumes that a static system model and learns a probabilistic model of the system. An unsuccessful execution would deviate from the model and the difference between two such traces represents the localization. Large distributed systems (open source systems eg., HDFS, HBase and commercial systems eg., Netflix, AWS) are complex and constantly evolving, invalidating this assumption. 
Spectroscope~\cite{SambasivanNSDI2011} assumes that a small number of unique execution paths exist in the system compared with the large number of underlying system traces, an assumption that does not hold for any but the smallest systems. GMTA~\cite{GuoFSE2020} makes two assumptions. First, it assumes that the model is known and traces can be accurately labelled based on the functionality they exercise. Second, it assumes that for each label, there exists a single canonical graph. The first assumption requires that the labelling be kept up-to-date with changing models but the second assumption only holds if there exists only a single execution path for given functionality, a premise that is untrue for large systems, as discussed at the beginning of the section. In summary, the simplifying assumptions made do not hold for large distributed systems.

In our work, we sidestep the problem of graph selection by considering \textit{sets} of traces rather than selecting a single pair of traces. From these, we derive \textit{aggregate} insights while preserving useful information for difference based diagnosis. Lightstep~\cite{Lightstep} represents the closest industry tool to \aname\ and addresses some of the same failure modes. Lightstep also compares traces in aggregate, but focuses on finding tags or markers in the traces containing errors. However, for failures in the recovery path, being able to identify that a call was not successful does not help with determining a mitigating action. In our example, Lightstep would follow the errors to the failed call from \textit{Payments service} to \textit{Token service}. This only represents one half of the localization and the incident could only be effectively mitigated by SREs understanding that the call to \textit{Token Fallback service} also failed in an attempt to recover from the failed call to \textit{Token service} - i.e. knowledge of both the missing call and the additional call. Furthermore, success or failure is an end to end property of a request and typically cannot be derived from a trace. For eg., a service returning an error in a user request does not necessarily imply a failed request; rather, it may be an indication to re-try the same.

In section \ref{design}, we describe our approach that compares sets of traces from steady-state operation and during the incident and analyzes their events and interactions to localize incidents.

\section{Design \& Methodology}
\label{design}

In our work, we use traces to localize incidents and thereby, speed up incident response. Localizing an incident highlights the absence (or presence) of event interactions during an incident that helps identify a location to apply a mitigation. 
Pairwise comparison is usually ineffective for localizing incidents since it produces false alarms, as described in Section~\ref{bkground_motivation}. 
As we now show, we can precisely localize incidents by comparing \textit{sets} of traces and using the \textit{structure} of traces to separate effects from their potential causes, retaining only the causes. 

\begin{figure}[t!]
\centering
\includegraphics[width=0.9\columnwidth]{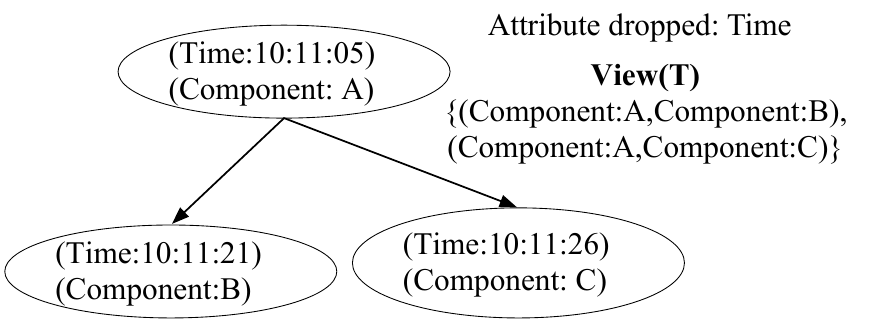}
\caption{Simple trace and an example of a view}
\label{ex_trace}
\end{figure}

We first describe View of a Trace which enables trace comparison. 
It is not possible to directly compare traces since individual traces include details such as timestamps that are different for every trace and IP addresses that are not necessarily consistent between any two traces. By dropping attributes that are not consistent across traces, views of traces make traces comparable. For example, to debug issues when a service is unable to talk to another, retaining service names is sufficient. If, instead, we would like to debug issues that impact a subset of service instances, retaining service instance names when generating a view of a given trace can be helpful. In this work, we only retain component names when generating views. We denote views by View(T) and refer to elements of a view as \textbf{ordered pairs}. In \Figure~\ref{ex_trace}, for example, the ordered pair (Component:A, Component:B) in View(T) corresponds to the edge ((Time:10:11:05, Component:A), (Time:10:11:21, Component:B)) in the trace, T.

\begin{figure*}[t!]
\centering
\includegraphics{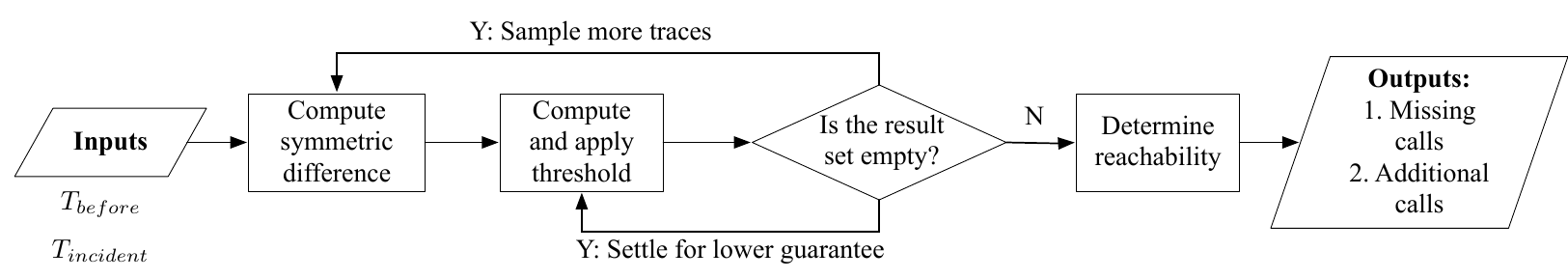}
\caption{\aname\ consists of applying three techniques: Symmetric difference, thresholding and reachability - in that order.}
\label{system_model}
\end{figure*}

\subsection{Inputs and Outputs}
Inputs to \aname\ consist of two sets of traces - traces drawn during the most recent steady-state operation of the system ($t_{before}$) and traces drawn during the incident ($t_{incident}$). These are sampled based on the incident start time specified by SREs. We expect the sampled traces to satisfy two constraints. 
First, the number of traces sampled in each of the two sets must be large enough that a majority of events or event interactions, if captured in underlying traces, are present in the sampled traces. Many large-scale systems generate millions~\cite{KaldorSOSP2017} of traces per day, but a much smaller sample size turns out to be sufficient for localization, as we will see in Section~\ref{guarantees}.

Second, traces are labeled as successful or unsuccessful based on an external success criterion. Examples of external criteria could include credit card charged in case of buying an item, the item displayed correctly when it is added to the product catalog, a HTTP status code of 200, an acknowledgement of data writes, etc. If a trace cannot be assigned a label, it is discarded (based on our data, less than 0.2\% of traces).

Outputs from our system should localize the incident under consideration rather than return the entire difference between the set of traces before and during the incident. SREs can investigate along two axes - a) Why are specific calls missing during the incident? and/or b) Why are other calls present \textit{only} during the incident? Based on what the investigation reveals, an appropriate mitigating action can be applied.

\subsection{System Overview}
\label{overview}
In \aname, we use aggregate information from witnessing a large set of traces and the causality of event interactions within individual requests to localize incidents. 
\Figure~\ref{system_model} shows the three techniques we use to localize an incident given sets of traces from before and during the incident.

The symmetric difference of $t_{before}$  and $t_{incident}$ is the set of ordered pairs that are in one of $\bigcup_{i=1}^{|t_{before}|}View(Trace_{i})$ or \newline $\bigcup_{j=1}^{|t_{incident}|}View(Trace_{j})$ but not both. If the changes produced by an incident are represented in $t_{incident}$ and at least one example of the correct interaction is in $t_{before}$, the symmetric difference \textit{will} contain the site where the mitigation is to be applied. To obtain a precise result, we employ thresholding and reachability.

We use thresholding to answer the question: Which ordered pairs in the symmetric difference are statistically significant and must be retained? 
Since $t_{before}$ and $t_{incident}$ are randomly sampled, one or more of the sampled traces may correspond to a code path that is rarely exercised. If such traces occur in one or the other set of traces, some ordered pairs will be part of the symmetric difference as a result of sampling randomness. The use of thresholding allows us to discard these. Using a threshold also addresses trace quality issues in individual traces that arise due to the best-effort nature of tracing.

After computing symmetric difference and applying the threshold, the result may still contain some superfluous ordered pairs. To understand how this may occur, assume two ordered pairs (a, b) and (b, c) are in the result. The edges in a trace represent event interactions. For a given trace, let's further assume that (a, b) and (b, c) correspond to edges ($e_1, e_2$) and ($e_2, e_3$) respectively. Reachability is the transitive closure of the edge relation of a graph. If we find that ($e_2, e_3$) is reachable from ($e_1, e_2$), we can discard the ordered pair (b, c) since its potential cause (a, b) is in the result. By establishing cause-and-effect relationships between edges corresponding to ordered pairs and eliminating the ordered pairs corresponding to effects, we use reachability to whittle down the result set for effective localization. Failure of a database call or third party vendor issues into which SREs have no visibility can be localized to a single leaf node or edge. For these, we expect to see effective localization even without the use of reachability. 

The three techniques build on each other - symmetric difference produces the initial result set while thresholding and reachability prune the result set such that the incident is effectively localized.

\subsubsection{Techniques:}
\label{techniques}
We describe in detail each of the techniques introduced in the previous section.
\heading{Symmetric Difference:}
To compute the symmetric difference, we only consider the successful executions in steady-state operation (unsuccessful requests in steady state could result from invalid credit card entry, insufficient stock, etc). $t_s$ represents successful executions in $t_{before}$ and we shorten $t_{incident}$ to $t_{inc}$ here. The result is the \textit{entire} set of changes between the two sets of traces. If calls made in traces of successful executions during the incident are in the symmetric difference, these could not possibly have been caused by the incident. Therefore, we remove them from our symmetric difference. Let $t_{inc_s}$ represent successful executions in $t_{inc}$. We obtain the symmetric difference as follows:
\begin{align*}
Missing \ Calls\ (M) &= \bigcup\limits_{i=1}^{|t_s|}View(t_{s(i)}) - \bigcup\limits_{j=1}^{|t_{inc}|}View(t_{inc(j)}) \\
Additional\ Calls\ (A) &= \bigcup\limits_{j=1}^{|t_{inc}|}View(t_{inc(j)}) - \bigcup\limits_{i=1}^{|t_{s}|}View(t_{s(i)}) \\
				& - \bigcup\limits_{k=1}^{|t_{inc_s}|}View(t_{inc_s(k)}) \\
D &= M \bigcup A \\
\end{align*}

\vspace{-0.75cm}
 \heading{Thresholding:}
We cannot use a flat threshold to determine the statistically significant ordered pairs because our threshold value can change not only as a result of system evolution but also based on the number of traces sampled. We derive our threshold as a function of frequency of calls in traces and the number of traces sampled. 
Frequency statistics can be computed in real time as traces are generated. Computed statistics can be stored in-memory since their memory footprint is small (order of a few hundred keys in a hash map). 

\begin{center}
Threshold, $t\ =\ N\ *(1 - e^{log(0.01)/n} ) $
\end{center}
We obtain $t$ by solving for $(1-p)^n < 0.01$, where $p$ is the probability that an ordered pair, $c$, occurs in a randomly sampled trace and is given by $\frac{t}{N}$. We assume that $c$ appears in $t$ traces, the size of the corpus from which frequency statistics are computed is $N$ and the number of sampled traces is $n$. The threshold, $t$, is such that if a call appears in more than $t$ of $N$ traces, there is a 99\% probability that at least one trace containing the call will be present in a sample of $n$ traces.

Given a threshold, if a call appears in more traces than the threshold and is unrelated to the incident, it will appear in both sets of traces with high probability and therefore not appear in the result. On the other hand, if the call is missing as a result of changes produced by an incident, evidence of the change will be seen in the sampled traces. Conversely, if the number of traces that a call occurs in is less than the threshold, it is discarded. SREs can choose a lower probability and re-compute the threshold for a less stringent guarantee.
\heading{Reachability:}
As discussed in Section \ref{overview}, we exploit reachability to achieve the minimal result set. To do so, we use the structure of individual traces. Given two ordered pairs and a trace, T, we first determine the possible edges that each ordered pair can correspond to. An ordered pair $o_{1}$ can correspond to many possible edges in a given trace since a view is generated by a lossy transform. Assume that ordered pairs $o_{1}$ and $o_{2}$ correspond to sets of edges represented by $s_{1}$ and $s_{2}$ respectively.  For example, given the ordered pair (Component:A, Component:B) and the trace from \Figure~\ref{ex_trace}, it would be mapped to a set containing the single edge - \{(Time:10:11:05, Component:A), (Time:10:11:21, Component:B)\}.

Next, we check if a cause-and-effect relationship exists between an edge in $s_1$ and one in $s_2$. If such a relationship is established, we discard the ordered pair corresponding to the effect while retaining its potential cause. We have reduced both the result set and the number of pairs to consider. We repeat this process for every pair of edges (that correspond to ordered pairs in the result set) in every trace until either we arrive at a single result or have explored all sampled traces.

Computing reachability is an expensive operation responsible for almost all of the time taken by \aname\ and is therefore applied after thresholding to reduce the number of ordered pairs to be considered. The time taken to establish reachability is O($|r|^2 * n$), where $|r|$ is the number of ordered pairs in the result set and n is the number of traces sampled. In the worst case, it will be necessary to consider edges in all sampled traces if none of the calls in the result set are related to others. In practice, many calls are related and time to establish reachability is much lower than the worst case bound.

\subsection{Application of \aname: An example}
\label{case_study_desc}
We now walk through an example of a simulated incident from the \company\ dataset which demonstrates how techniques in \aname\ apply end to end and highlights trade-offs SREs often need to make when an incident produces changes in a small number of traces. We simulate interruption in communication between PaymentService and OrderMgmtService. For users purchasing items, this call is required to validate the purchase. Interruption results in users being unable to place orders. Therefore, we want to highlight the missing call from PaymentService to OrderMgmtService. Assume that the probabilistic guarantee is 0.99 (if a call appears in more traces than the threshold, it is in the sampled traces with 99\% probability) and $t_{before}$ and $t_{incident}$ each contain 2K traces. 

\aname\ computes a set of results, the elements of which are ordered pairs. The symmetric difference produces a result set of size 21 but after applying thresholding, the result set is empty. This implies that the sampled traces do not contain evidence of the correct execution, the changes produced by the incident, or both. SREs can now take two actions:

\heading{Reduce the threshold:} An SRE may decide to trade-off number of results for time i.e. it is acceptable if the computed result has some irrelevant elements. The SRE will now choose a lower probability and re-compute the threshold. In our example, the SRE chooses to drop the probability to 0.75. The result from symmetric difference contains 21 ordered pairs. The size of the result is now 11 after applying the new threshold. On applying reachability, we obtain a result of size 2 - the expected result and an additional, irrelevant suggestion.

\heading{Sample more traces:} An SRE can also decide to trade-off time for number of results i.e. additional time is acceptable for fewer (ideally zero) irrelevant results. Since this is a simulated incident, we know that we can obtain the expected result with high probability by sampling 4K traces in each set. With the resampled traces, we compute the symmetric difference (result size is 41) and apply thresholding (result size reduced to 3). Applying reachability yields exactly the expected result. 

The choice to trade-off time or number of results is situational - for example, if trading off number of results for time produces many false positives, SREs may pivot and sample more traces instead. 
\section{Evaluation}
\label{eval}
In Section \ref{guarantees}, we discuss how the initial sample size is determined and used. To evaluate \aname, we simulate incidents based on how we expect traces to change for each incident category. Section \ref{exp_method} makes the case for simulating incidents and we discuss \textit{how} we mutate traces. Section \ref{baselines} describes the baselines we compare against. Finally, we compare the results of \aname\ with baseline techniques employed in prior work. We answer the following questions:
\begin{enumerate}
\itemsep0em
\item How is the initial sample size determined? (Section \ref{guarantees})
\item How do the results produced by \aname\ compare with baseline techniques? How do the individual techniques in \aname\ impact the result produced? (Section \ref{pre_acc_desc})
\item How does the time to result compare with baseline techniques? (Section \ref{time})
\end{enumerate}
Finally, in Section \ref{impl_deet}, we present some highlights of integrating \aname\ with Jaeger~\cite{Jaeger} for online comparison of traces.

\subsection{Determining the initial sample size}  
\label{guarantees}
We discuss how we use \aname's probabilistic guarantees to determine the initial sample size from underlying traces and frequency statistics. This serves as an input when sampling $t_{before}$ and $t_{incident}$ for localization.

A structural change to a trace consists of ordered pairs that are missing during an incident which would normally be present in traces during steady state operation or additional calls only occur during an incident. From our discussion of thresholding in Section \ref{techniques}, we provide a probabilistic guarantee that evidence of the correct interaction as well as structural changes to traces are represented in $t_{before}$ and $t_{incident}$ respectively if they appear in more traces than the threshold.

\begin{figure*}[t!]
\centering
\begin{subfigure}{0.32\textwidth}
\includegraphics[width=\linewidth]{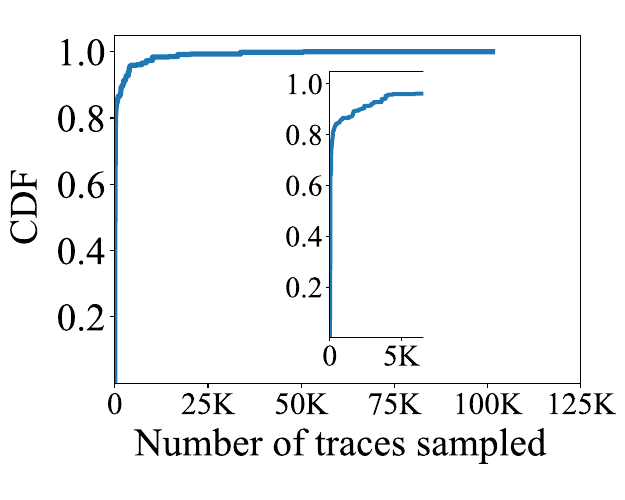}
\caption{DSB (22K traces)}
\end{subfigure}\enskip
\begin{subfigure}{0.32\textwidth}
\includegraphics[width=\linewidth]{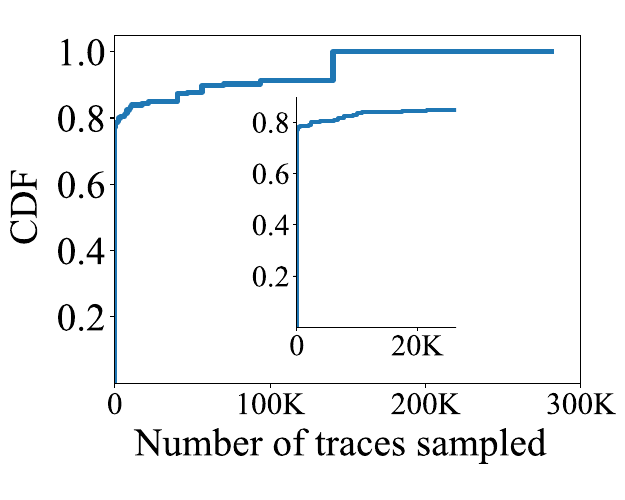}
\caption{HDFS (60K traces)}
\end{subfigure}\enskip
\begin{subfigure}{0.32\textwidth}
\includegraphics[width=\linewidth]{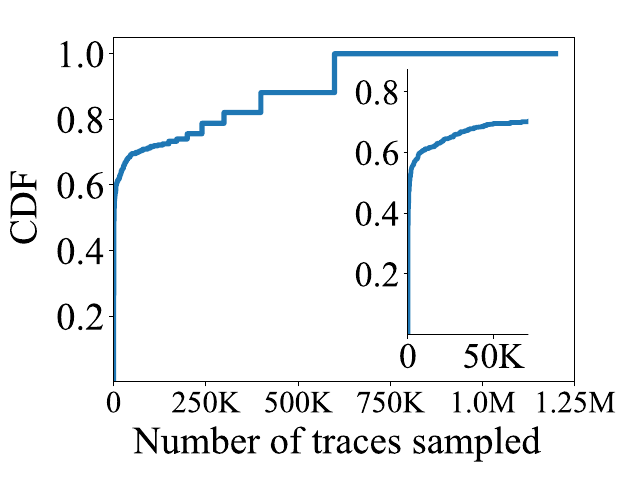}
\caption{\company (250000+ traces)}
\end{subfigure}\enskip
\caption{CDF of the number of traces to be sampled to identify any possible missing edge. The inlaid snippet of the CDF shows that a majority of calls can be identified with a sample that is orders of magnitude smaller.}
\label{cdf_plots}
\end{figure*}

We can plot a Cumulative Distribution Function (CDF) of the percentage of ordered pairs probabilistically guaranteed to be represented for a given number of sampled traces. \Figure~\ref{cdf_plots} depicts these for our three datasets. From the CDFs, we observe that although a very large number of traces would need to be sampled to identify every possible call (if it were missing), we find that a majority of calls can be identified with a sample that is orders of magnitude smaller. Accordingly, we sample 4K (of 20K) traces for DSB, 8K (of 60K) for HDFS and 20K (of 250K+) for \company\ in our experiments.

\subsection{Experimental Methodology}
\label{exp_method}
\begin{table*}[t]
\small
\centering
\caption{This table explains how we simulate the three failure modes we consider. For each, we describe the input, how traces are mutated and the expected output. We also specify the conditions that need to be satisfied in each case for a trace to be mutated. All mutated traces represent unsuccessful executions.}
\label{simulator}
\begin{tabular}{ |>{\raggedright\arraybackslash}p{3.2cm}|>{\raggedright\arraybackslash}p{2.3cm}|>{\raggedright\arraybackslash}p{2.5cm}|>{\raggedright\arraybackslash}p{4cm}|>{\raggedright\arraybackslash}p{3cm}|}
\hline
\textbf{Incident Category} & \textbf{Input} & \textbf{Condition for mutation} & \textbf{How are traces mutated?} & \textbf{Expected Result} \\
\hline
Component down & Randomly chosen component & Vertex corresponding to component is present in trace & Delete all edges to vertices corresponding to chosen component as well as the subgraph beneath each edge & Component chosen as input\\
\hline
Component Unreachable & Randomly chosen ordered pair &  At least one edge corresponding to ordered pair is present in trace & Delete all edges corresponding to the chosen ordered pair as well as the subgraph beneath each edge & Ordered pair chosen as input \\
\hline
Buggy failure recovery & Randomly chosen ordered pair & At least one edge corresponding to ordered pair is present in trace & Delete all edges corresponding to the chosen ordered pair as well as the subgraph beneath each edge, then add an edge at each call site representing an attempt to recover from failure & Ordered pair chosen as input and additional ordered pair attempting recovery\\
\hline
\end{tabular}
\end{table*}

\begin{table*}[t!]
\centering
\caption{\aname\ computes exactly the expected result for all but a few cases. In contrast, NodeCount and EdgeCount produce wrong answers for 30-50\% of simulations. Answer = Set of localizations returned, Exact Answer = Answer is minimal, Superfluous Answer =  Answer subsumes expected result, Wrong Answer = Answer does not contain expected result,  No Answer = Answer is the null set.}
\label{results_desc}
\begin{tabular}{ |p{0.1\linewidth}|p{0.1\linewidth}|p{0.1\linewidth}|p{0.1\linewidth}|p{0.1\linewidth}|p{0.09\linewidth}|p{0.1\linewidth}|} 
 \hline
 & & 
 \textbf{Number of simulations} & 
\textbf{Exact \newline Answer(\%)} &
\textbf{Superfluous \newline Answer(\%)} &
\textbf{Wrong \newline Answer(\%)}  &
\textbf{No \newline Answer(\%)} \\
 \hline
\multirow{3}{=}{\textbf{\aname\ } } & DSB & 602 & 99.83 (601) & 0.17 (1) & 0 & 0 \\\cline{2-7}
			        & HDFS & 401& 98.50 (395) & 0.25 (1) & 1.25 (5) & 0 \\\cline{2-7}
			        & \company\ & 418 & 99.76 (417) & 0.24 (1) & 0 & 0 \\
\hline
\hline
\multirow{3}{=}{\textbf{NodeCount}} & DSB & 602 & 52.82 (318) & 7.8 (47) & 37.21 (224) & 2.16 (13) \\\cline{2-7}
			        & HDFS & 401 & 21.95 (88) & 29.68 (119) & 47.63 (191) & 0.75 (3) \\\cline{2-7}
			        & \company\ & 418 & 25.11 (105) & 21.77 (91) & 48.80 (204) & 4.41 (18) \\
\hline
\hline
\multirow{3}{=}{\textbf{EdgeCount}} & DSB & 602 & 58.47 (352) & 2.66 (16) & 36.38 (219) & 2.49 (15) \\\cline{2-7}
			        & HDFS & 401 & 63.59 (255) & 4.49 (18) & 31.17 (125) & 0.75 (3) \\\cline{2-7}
			        & \company\ & 418 & 31.81 (133) & 16.27 (68) & 44.50 (186) & 7.42 (31) \\
\hline
\end{tabular}
\end{table*}

\begin{figure*}[th!]
\centering
\begin{subfigure}{0.3\linewidth}
\includegraphics[width=\linewidth]{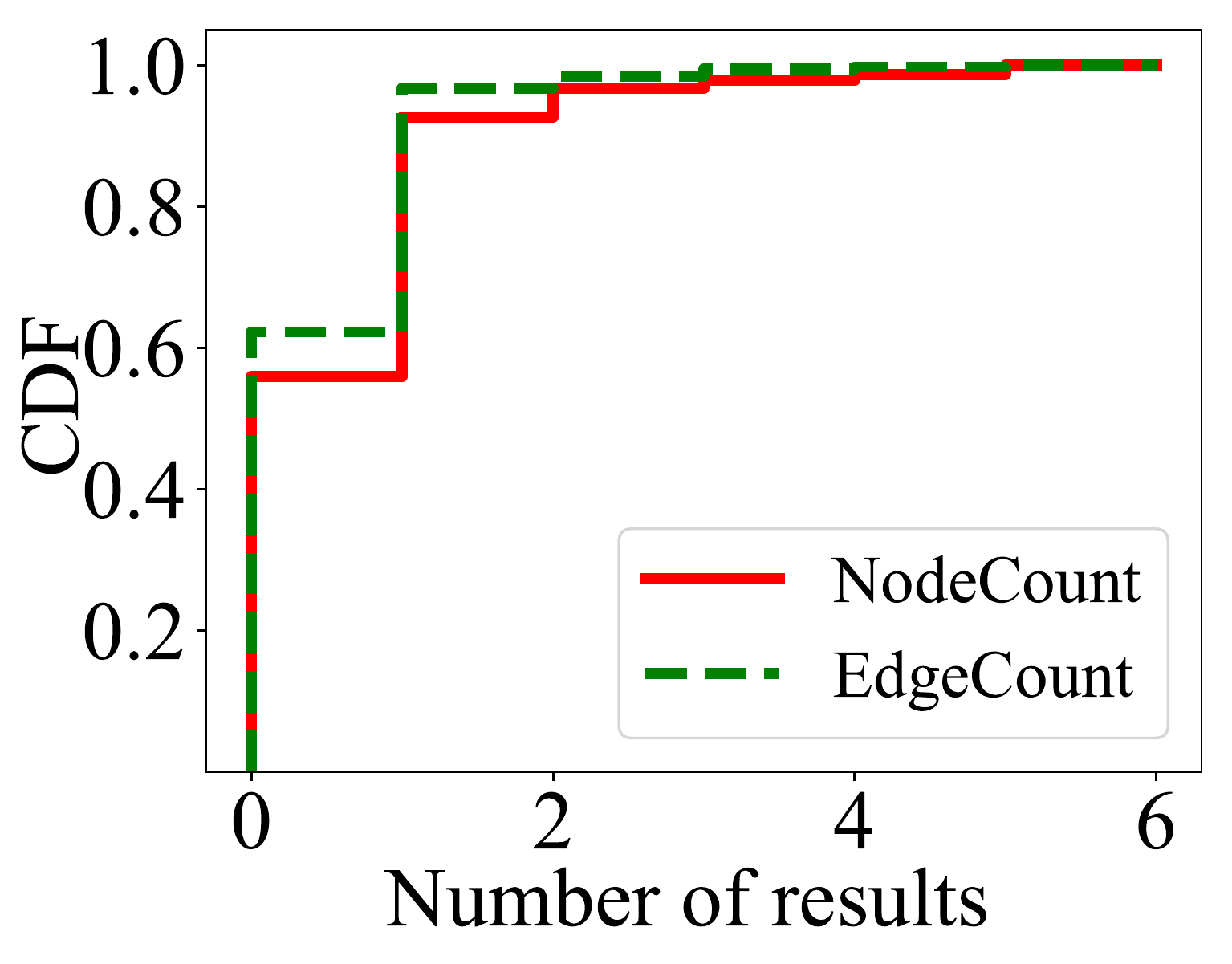}
\caption{DSB}
\end{subfigure}\enskip
\begin{subfigure}{0.3\linewidth}
\includegraphics[width=\linewidth]{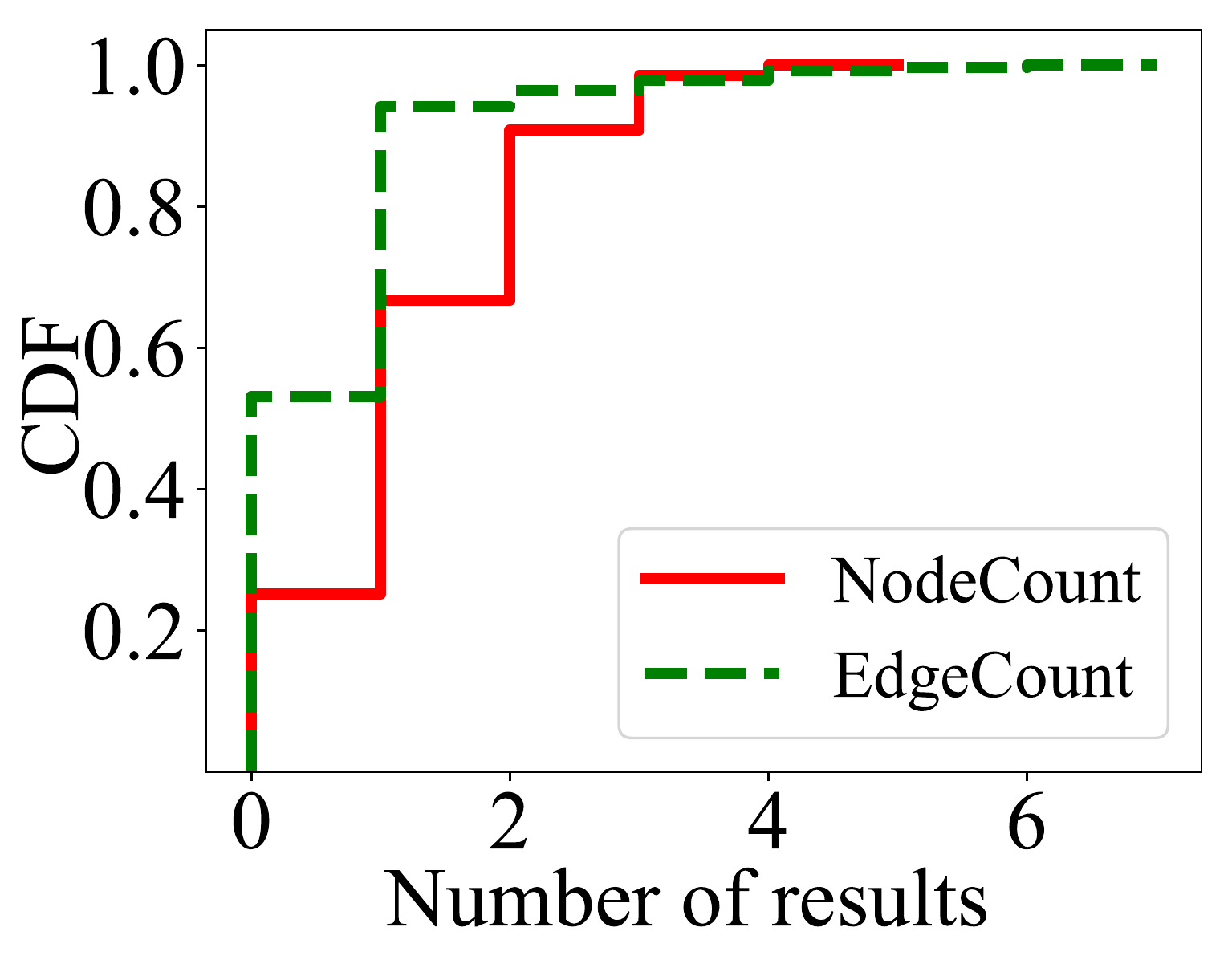}
\caption{HDFS}
\end{subfigure}\enskip
\begin{subfigure}{0.3\linewidth}
\includegraphics[width=\linewidth]{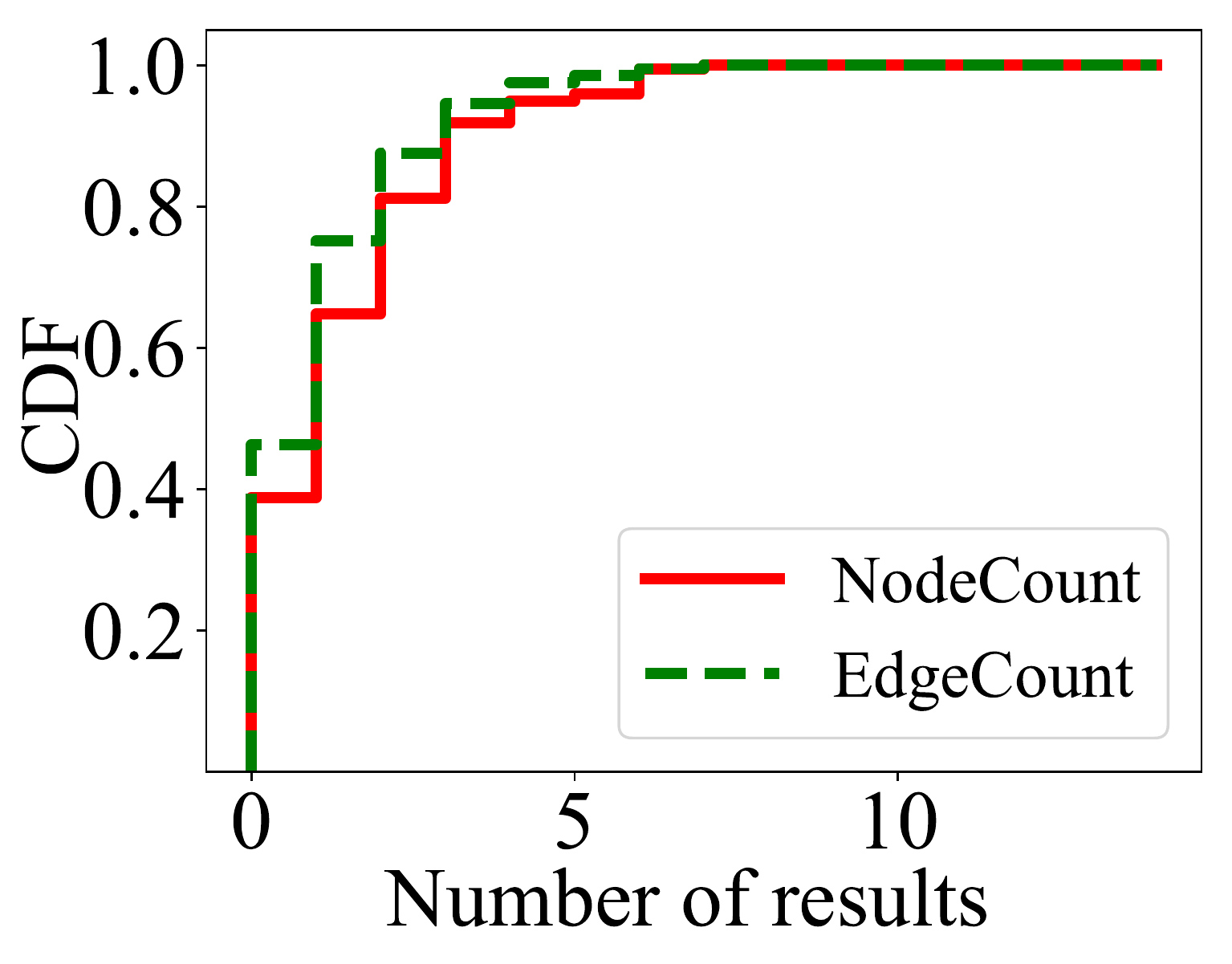}
\caption{\company}
\end{subfigure}
\caption{For the cases when NodeCount and EdgeCount produce results, we plotted a CDF of number of results. \aname, meanwhile, produces exactly the expected answer for all of these cases.}
\label{pre_cdf}
\end{figure*}

\begin{figure*}[h!]
\centering
\begin{subfigure}{0.3\linewidth}
\includegraphics[width=\linewidth]{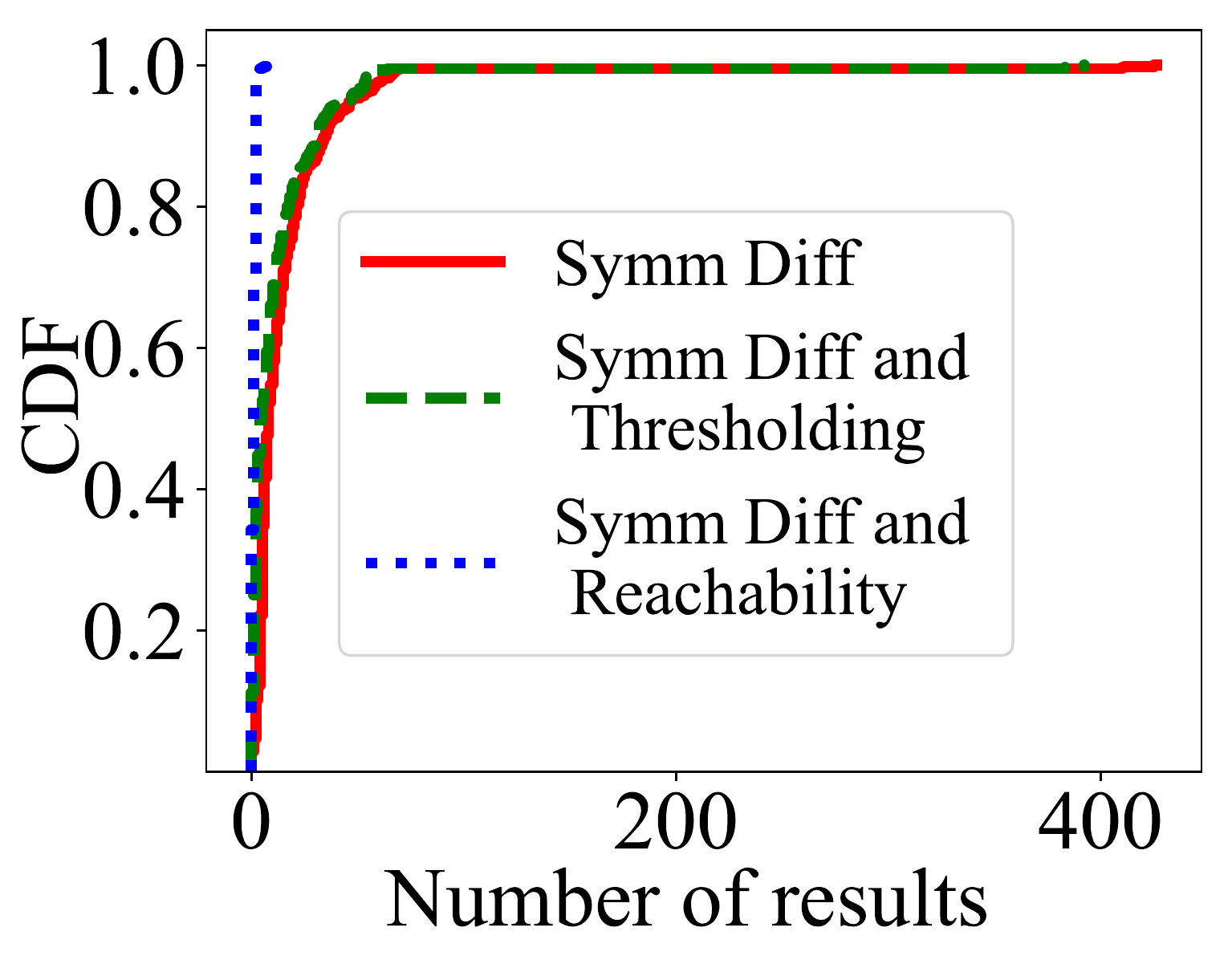}
\caption{DSB}
\end{subfigure}\enskip
\begin{subfigure}{0.3\linewidth}
\includegraphics[width=\linewidth]{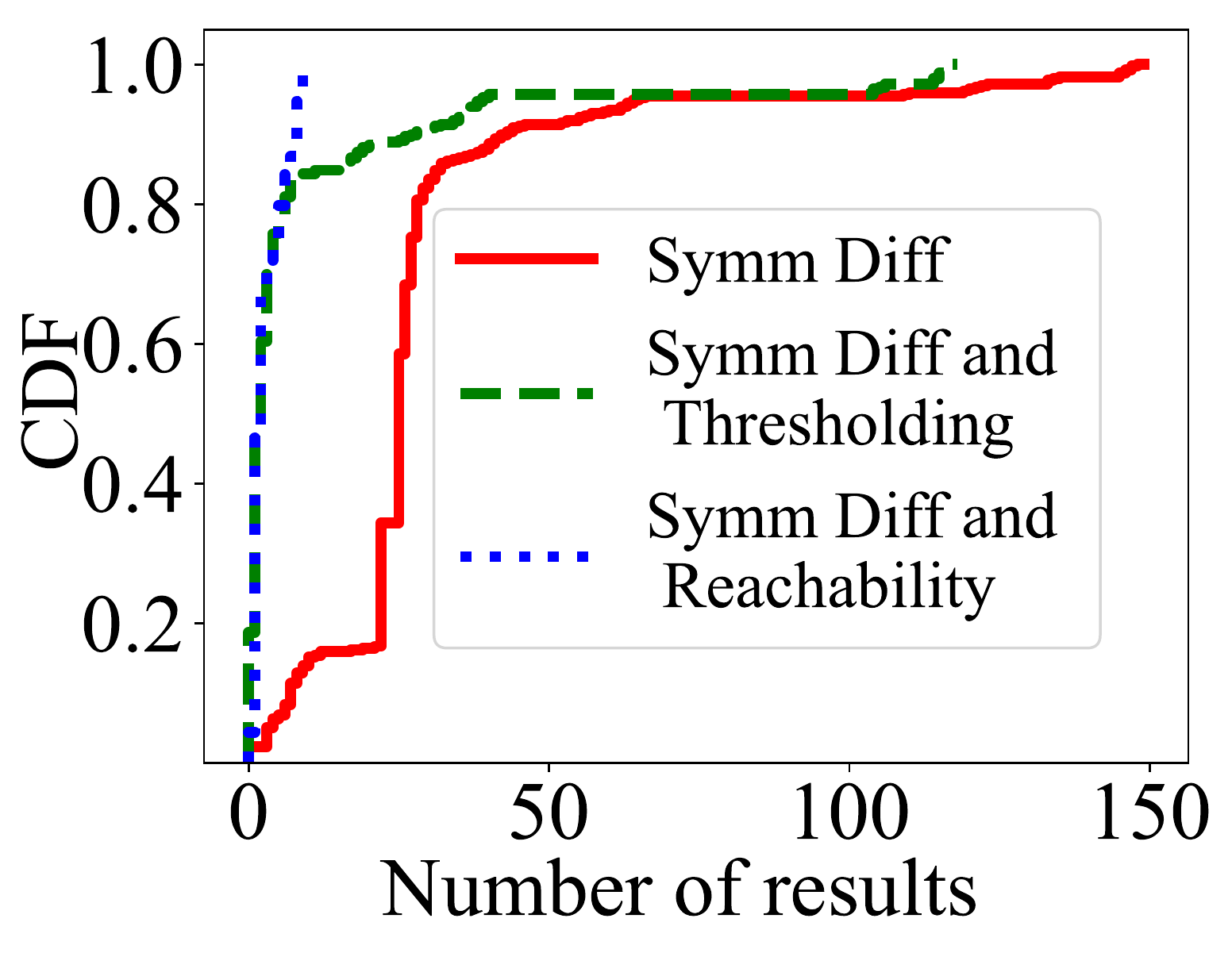}
\caption{HDFS}
\end{subfigure}\enskip
\begin{subfigure}{0.3\linewidth}
\includegraphics[width=\linewidth]{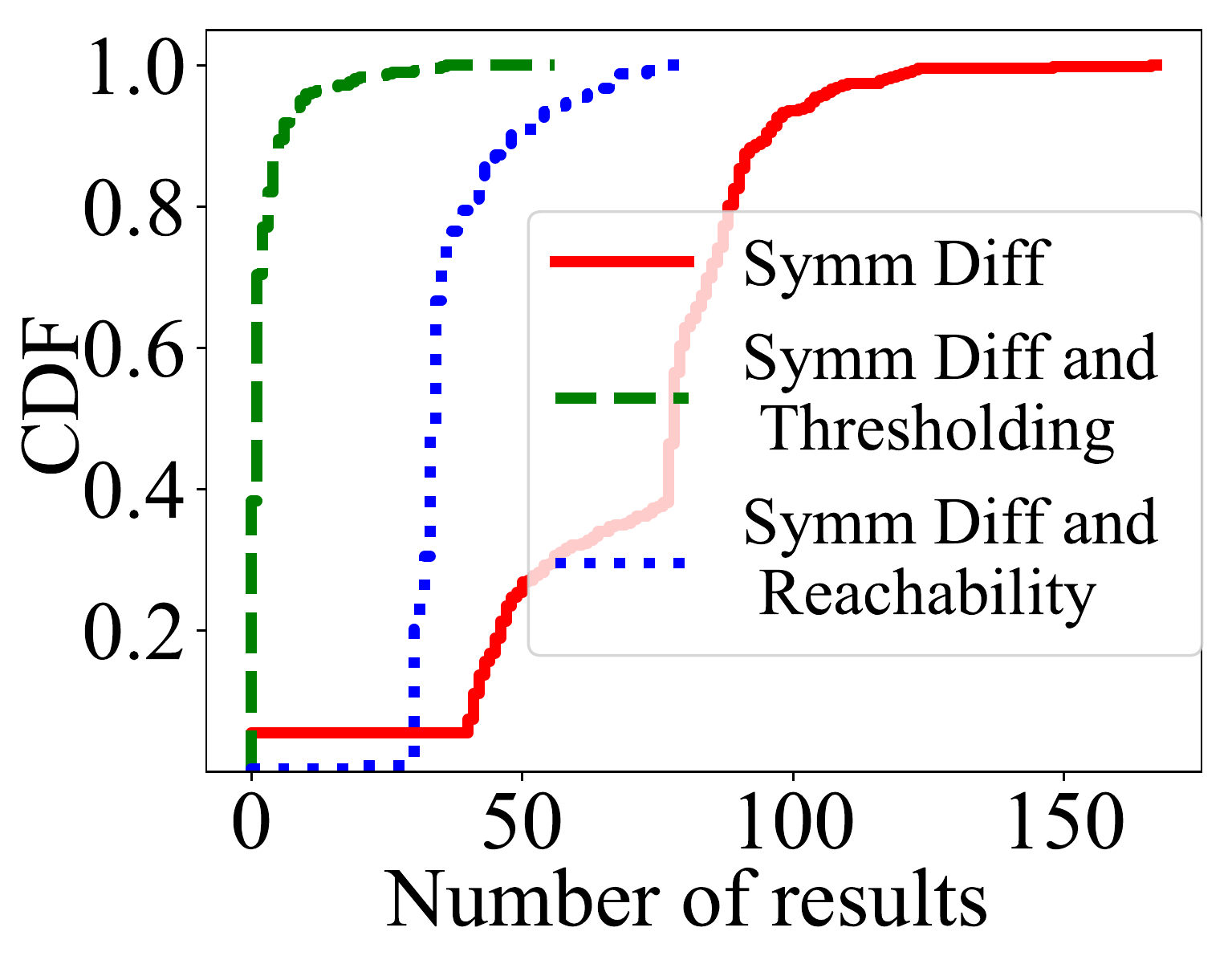}
\caption{\company}
\end{subfigure}
\caption{CDFs of the number of results returned when we apply one or two techniques. Since the \company\ dataset is noisy, symmetric difference and thresholding performs best, while symmetric difference and reachability generate the best results for DSB and HDFS. When all three techniques of \aname\ are applied, the result is obtained is exactly the mitigation site.}
\label{impact_techniques}
\end{figure*}

\begin{figure*}[h!]
\centering
\begin{subfigure}{0.3\linewidth}
\includegraphics[width=\linewidth]{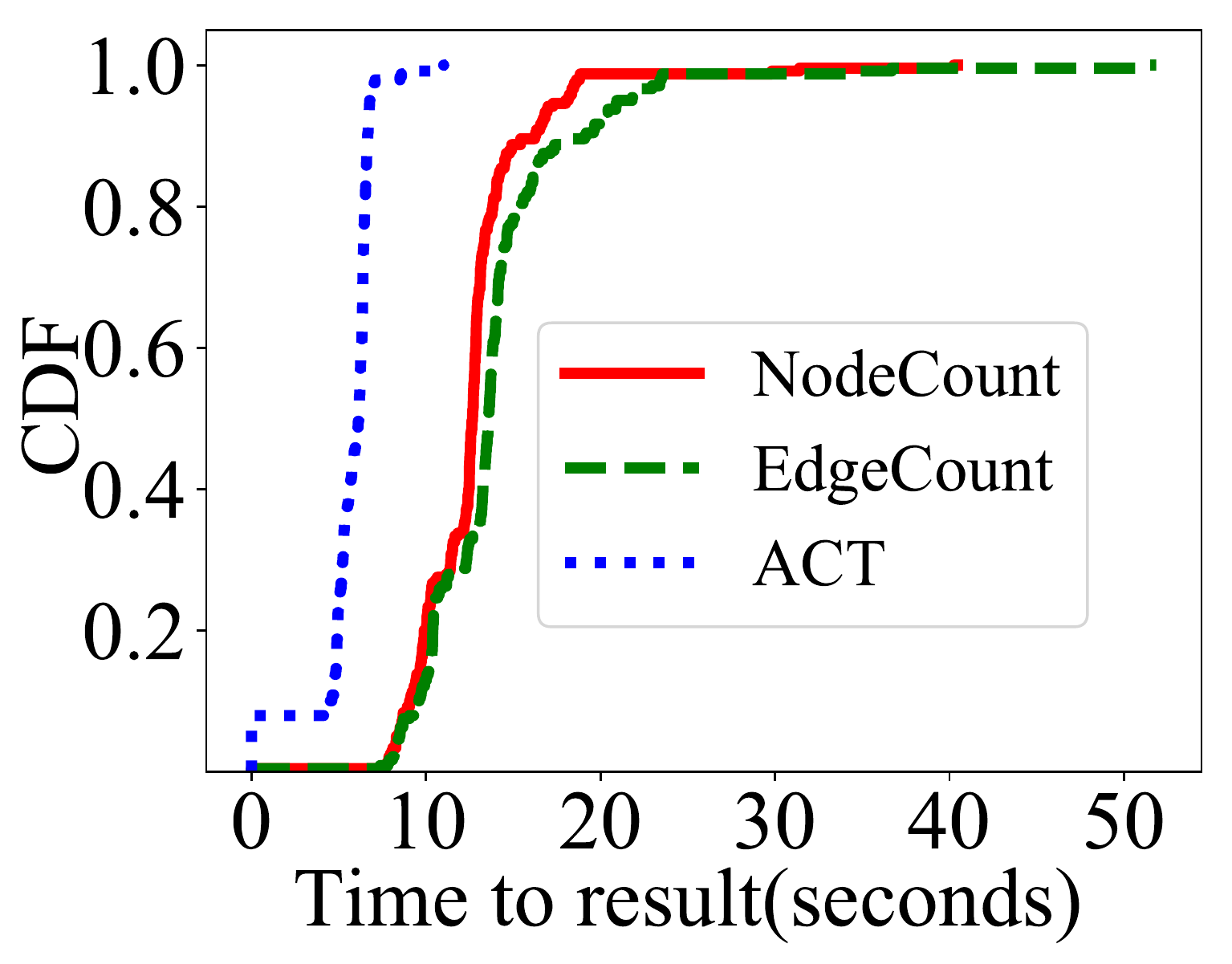}
\caption{DSB}
\end{subfigure}\enskip
\begin{subfigure}{0.3\linewidth}
\includegraphics[width=\linewidth]{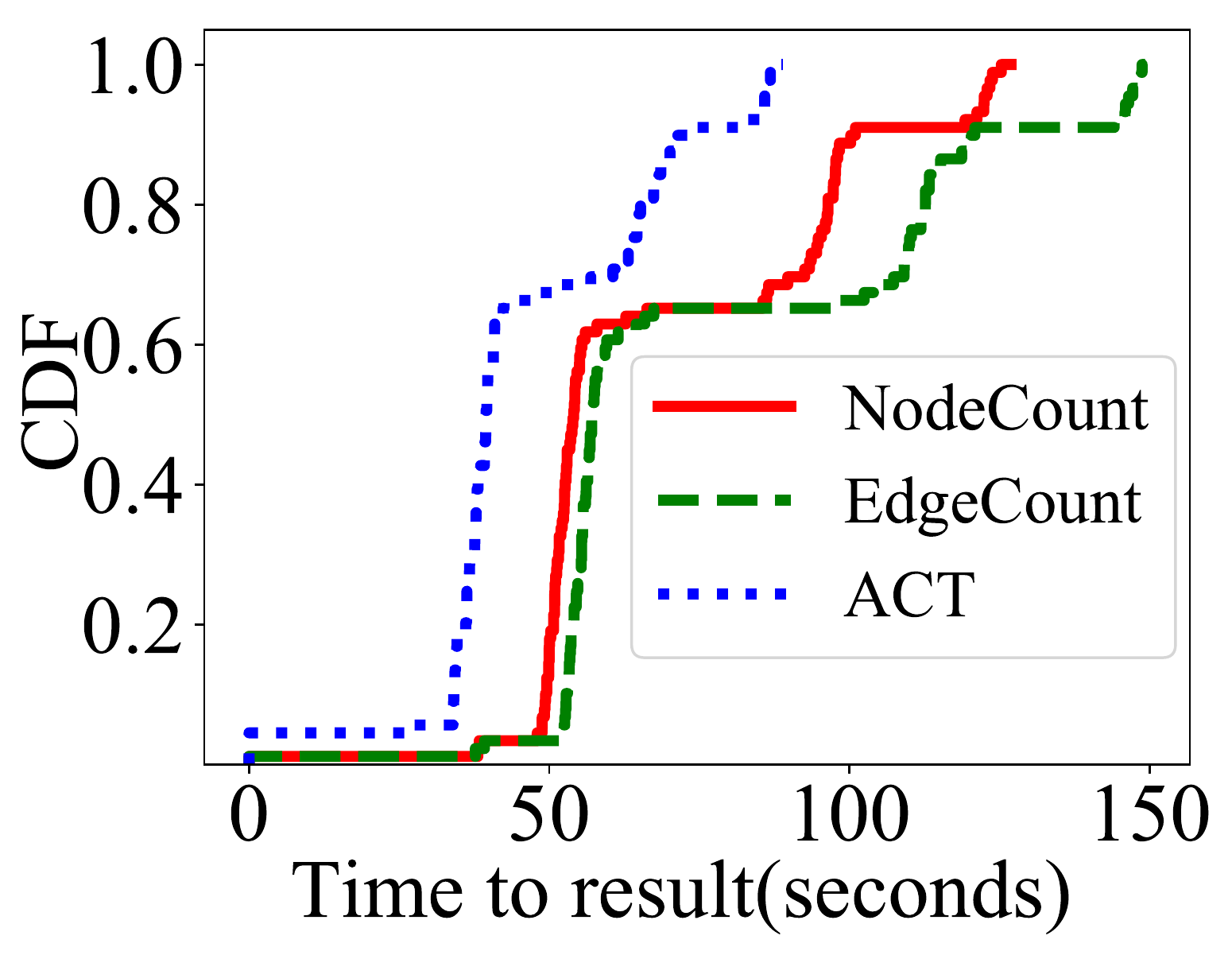}
\caption{HDFS}
\end{subfigure}\enskip
\begin{subfigure}{0.3\linewidth}
\includegraphics[width=\linewidth]{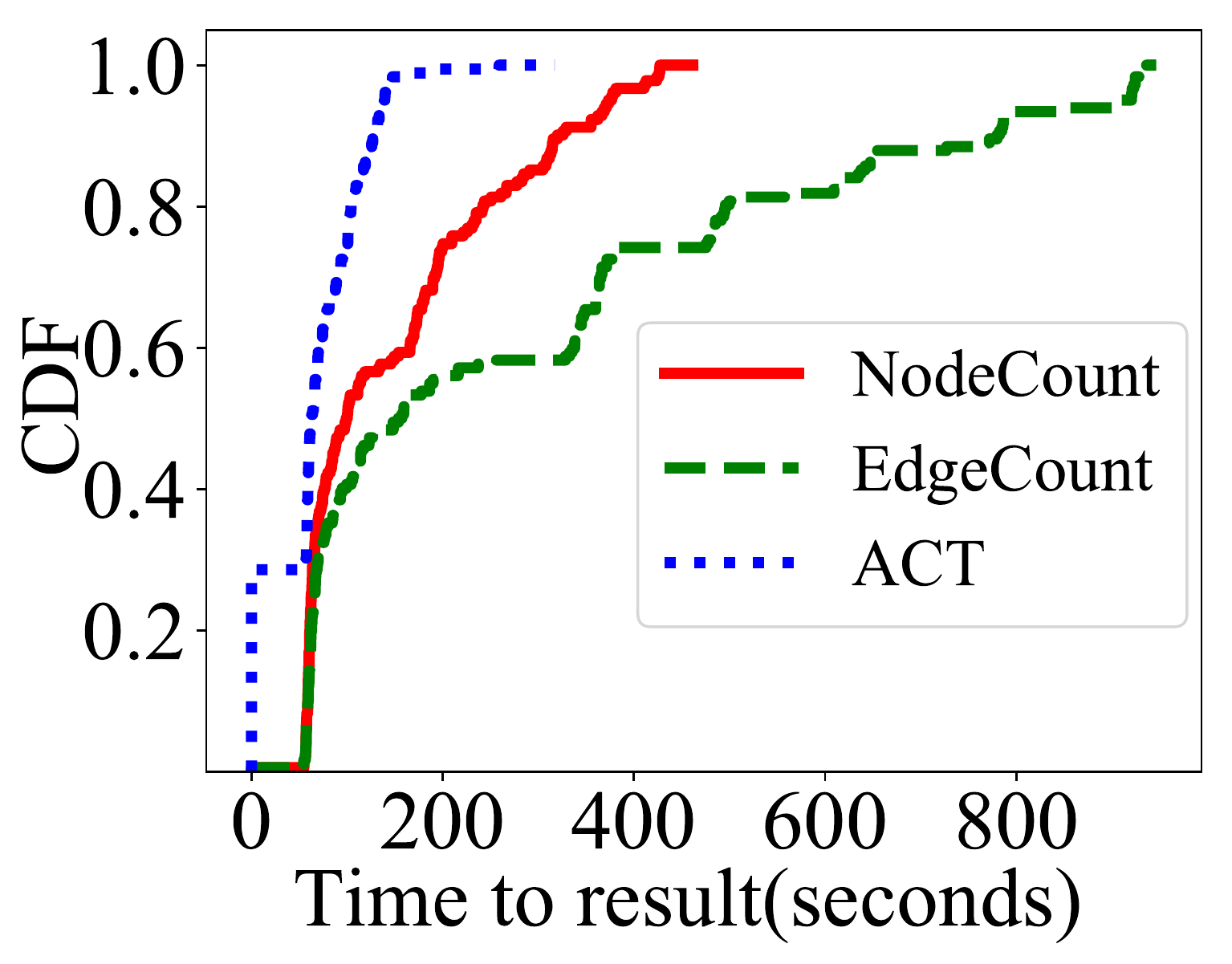}
\caption{\company}
\end{subfigure}
\caption{CDFs of time taken to obtain a result. Reachability accounts for most of the time taken by \aname\ . Nodecount and Edgecount have highly variable time to result since trace of every unsuccessful execution needs to be compared with that of every successful execution and the number of unsuccessful executions can vary widely.}
\label{time_taken}
\end{figure*}

To conduct a quantitative evaluation of \aname\ using data from real incidents, we would have needed to collect traces during steady-state operation and then again when incidents occur. Although a large number of incidents occur (anecdotally, three or four every day), we are only interested in those in one of the categories described. Identifying these and capturing traces while they are still retained remains a challenge. 

From our incident study and observations of traces generated when we inject faults, we have a good grasp on how we expect the structure of traces to change for each incident category. Therefore, simulating incidents can serve as a good proxy. Simulation not only allows us to apply \aname\ to a wide range of scenarios but is also useful in testing its limits.

To simulate an incident, we randomly sample two sets of traces which we designate as $t_{before}$ and $t_{incident}$ respectively. For each incident category, Table \ref{simulator} describes inputs, \textit{how} traces are mutated and expected output. Some fraction of traces in $t_{incident}$ that satisfy the condition for mutation are mutated to represent traces that would have been generated during the incident being simulated, while traces in $t_{before}$ remain unmodified. All mutated traces represent unsuccessful executions. An unsuccessful execution is one for which we evaluate some external criteria and determine that the user request corresponding to the execution did not succeed. We choose simulations uniformly at random. $t_{before}$ and $t_{incident}$ serve as inputs to the different techniques.

We use three trace datasets in our evaluation. These consist of a production dataset from \company\ and two open-source datasets - DeathStar Benchmark (DSB)~\cite{GanASPLOS2019}, a micro-services benchmark and Hadoop Distributed File System (HDFS)~\cite{ShvachkoMSST2010} traces. \company\ has about 4500-5000 services, the dataset captures user requests as they purchase items during a week in November 2019. User requests to start a session and complete a purchase account for nearly two thirds of the requests; the remaining third is distributed across twenty other request types that span different system functions. Examples include changing user address and payment modes as well as updating items or item quantities. The captured requests record 250+ unique services and databases and 850+ unique calls. Vertices and edges represent services and calls between services respectively. DSB traces were generated by deploying the benchmark on a single machine and capturing traces of different API types. HDFS traces were generated by deploying HDFS on a 9-node cluster and consists of traces obtained by reading and writing files of various sizes. The DSB and HDFS traces are in X-Trace~\cite{FonsecaNSDI2007} format and are captured at a lower level of abstraction where vertices represent execution of lines of code and edges represent the execution flow.

\subsection{Baseline techniques}
\label{baselines}
In prior work, graph analysis approaches~\cite{SambasivanNSDI2011, ChenDSN2002} transform graphs into vectors (by counting nodes or edges or converting them into strings) and compare pairs of graphs. The result returned is a pair of $ \langle Successful, Unsuccessful \rangle$ traces such that they exercise the same execution path and are separated by the shortest distance. "Shortest" is precisely defined based on the distance metric and the representation used. NodeCount and EdgeCount represent traces as vectors containing the counts of components and calls between components respectively and use $L_2$ distance as the distance metric. Spectroscope~\cite{SambasivanNSDI2011} linearizes traces to produce an event string and uses string edit distance as its distance metric. 

Since graph selection is not viable, we have adapted the different techniques to return the best result after comparing \textit{all} pairs of traces. The inputs are vectors or string representations of traces in $t_{before}$ and $t_{incident}$. For the resultant pair of traces, we compute the symmetric difference of the view of traces and apply reachability. This final step focusses attention on only the relevant results and is not employed in prior work. We take this step to be able to compare the results from the baselines with \aname. A single experiment comparing linearized traces took multiple \textit{hours} as compared to the few seconds taken by other techniques. Hence, we ran simulations comparing \aname\ with NodeCount and EdgeCount only.

\subsection{Results}
\heading{Result Quality}
\label{pre_acc_desc}
Table \ref{results_desc} summarizes the results for the simulations for which the change produced by the simulation is reflected in the sampled traces. We conducted hundreds of simulations for each dataset with the number of simulations for each incident category being roughly equivalent. As can be observed, \aname\ computes exactly the expected result for all but a few cases. In contrast, NodeCount and EdgeCount compute irrelevant results for 30-50\% of simulations for which the changes are in the sampled traces.

Additionally, when NodeCount and EdgeCount produce the expected result (in 2.5\% to 30\% of the scenarios) depending on the technique and dataset, results include false positives. From our experiments, EdgeCount produces false positives in fewer scenarios than NodeCount. \Figure~\ref{pre_cdf} shows the CDF of the number of results produced by NodeCount and EdgeCount. 

\heading{Impact of individual techniques}
To measure the impact of thresholding and reachability, we consider simulations for which \aname\ returns exactly the expected result, since the effects can be most clearly seen for these simulations. For the selected simulations, we employ combinations of one or two techniques and re-run them. \Figure~\ref{impact_techniques} visualizes the results we obtain. It is immediately apparent that computing symmetric difference with thresholding produces the best results for the \company\ dataset indicating a noisier dataset than HDFS or DSB. Reachability plays a bigger role for DSB and HDFS datasets since these have more depth as compared with \company\ dataset, in which graphs are wide and shallow. Across the board, the three techniques taken together are more powerful than any pair of techniques. 

\heading{Time taken to obtain result}
\label{time}
\Figure~\ref{time_taken} represents CDFs of time taken when the most number of traces are sampled for each dataset. Reachability computations account for almost all of the time taken by \aname. Symmetric difference and thresholding reduce the number of pairs for which reachability computations need to be performed - the time for which is negligible in comparison to reachability computation. \aname\ has a time bound of O($|r|^2*n$), which is linear in the number of sampled traces, as discussed previously. The baseline techniques, however, have a time bound of O(s*u), where s is the number of successful executions in $t_{before}$ and u is the number of unsuccessful executions in $t_{incident}$. Since the trace of every unsuccessful execution is compared with the trace of every successful execution, the time taken is quadratic.

\subsection{Integration with Jaeger: Implementation details}
\label{impl_deet}
We have integrated our approach with Jaeger~\cite{Jaeger} to enable online comparison of sets of traces. Jaeger is an open source, end-to-end distributed tracing tool that enables monitoring and troubleshooting complex distributed systems. It currently provides a feature that allows users to select and compare a pair of traces. The obvious drawback is that users need to know \textit{which} traces to compare. We have extended the UI to compare sets of traces instead. Rather than requiring users to select traces as input, we accept as input the time since the incident started. This enables us to split the traces into before and after sets. For the purposes of our integration, we use HTTP status codes in the traces to mark them as successful or unsuccessful - a trace with any span returning a non-zero status code is considered unsuccessful. In general, SREs can use any criteria to label traces as successful or unsuccessful. With the two sets of traces and their labels as inputs, we extended Jaeger-UI to implement and visualize \aname.

\section{Related Work}
\label{related_work}
Section \ref{bkground_motivation} contrasts \aname\ with much of the relevant prior work. We reiterate how \aname\ differs from closest related work here.

\heading{Approaches based on trace analysis} Prior works (~\cite{BeschastnikhCACM2016, GuoFSE2020, SambasivanNSDI2011, ChenDSN2002, BarhamHotOS2003}) employ pairwise comparison techniques. Since graph selection based only on the structure of the traces is not viable for large distributed systems, as argued in Section \ref{bkground_motivation}, these are better suited for interactive debugging. For automated incident localization, \aname\ avoids the problem of graph selection by comparing \textit{sets} of traces. 

\heading{Tools in industry} Various tools have been built to derive value from trace data in industry~\cite{DanLuu, Edgar, Slack, Cindy, Lightstep}. Lightstep~\cite{Lightstep} in particular takes a similar approach to ours in that it analyzes traces in aggregate by comparing traces from different time periods, but differs in that it attempts to find tags that characterize traces which contain erroneous spans. By identifying missing as well as additional calls via comparing sets of traces, \aname\ addresses the limitations of Lightstep's follow-the-errors approach, discussed at the end of Section \ref{bkground_motivation}. 
\section{Conclusion}
\label{summary}
\aname\ combines the use of aggregate and causal information in traces to effectively localize incidents. \aname\ identifies \textit{exactly} the mitigation site in all but a few cases. While witnessing a large number of traces is necessary to derive aggregate insights from traces, for a majority of incidents that produce structural changes in traces, the number of traces to be sampled is orders of magnitude smaller than the underlying traces captured by the system, making it viable for use in reducing the time to localize and thereby, resolve incidents.


\bibliographystyle{unsrt}		
\bibliography{ref}

\end{document}